\documentclass[12pt]{amsart}
\usepackage[headings]{fullpage}

\usepackage{amsmath,amssymb,amsthm}
\usepackage{graphicx}
\usepackage{enumerate}
\usepackage{url}
\usepackage{slashed}
\usepackage{bm}
\usepackage[german,english]{babel}

\usepackage[bookmarks=true,%
    colorlinks=true,%
    linkcolor=blue,%
    citecolor=blue,%
    filecolor=blue,%
    menucolor=blue,%
    urlcolor=blue,%
    breaklinks=true]{hyperref}

\usepackage[all]{xy}
\usepackage{verbatim}
\usepackage{tikz-cd}

\newtheorem{thm}{Theorem}[]
\newtheorem*{thm*}{Theorem}

\newtheorem{ex}[thm]{Example}

\newtheorem{defn}[thm]{Definition}
\newcommand{\param}{{\mathchoice{\mkern1mu\mbox{\raise2.2pt\hbox{$
\centerdot$}}
\mkern1mu}{\mkern1mu\mbox{\raise2.2pt\hbox{$\centerdot$}}\mkern1mu}{
\mkern1.5mu\centerdot\mkern1.5mu}{\mkern1.5mu\centerdot\mkern1.5mu}}}
\newcommand{\thereals}{\ensuremath{\mathbb{R}}}

\newcommand{\manifold}[1]{\ensuremath{\mathcal{#1}}}

\newcommand{\nbhd}[1]{\ensuremath{\mathcal{#1}}}

\newcommand{\bphim}{\ensuremath{\partial_{\phi}\manifold{M}}}
\newcommand{\phim}{\ensuremath{\phi(\manifold{M})}}
\newcommand{\mhat}{\ensuremath{\widehat{\manifold{M}}}}

\newcommand{\aboundary}{\textit{a}-boundary}

\renewcommand{\setminus}{{\smallsetminus}}

\begin{document}

\title{Examples of covering properties of boundary points of space-times}
\author{Ingrid Irmer}
\address{Mathematics Department\\
Technion, Israel Institute of Technology\\
Haifa, 32000\\
Israel}
\email{ingridi@technion.ac.il}

\begin{abstract}
The problem of classifying boundary points of space-time, for example singularities, regular points and points at infinity, is an unexpectedly subtle one. Due to the fact that whether or not two boundary points are identified or even ``nearby'' is dependant on the way the space-time is embedded, difficulties occur when singularities are thought of as an inherently local aspect of a space-time, as an analogy with electromagnetism would imply. The completion of a manifold with respect to a pseudo-Riemannian metric can be defined intrinsically, \cite{aboundary}. This is done via an equivalence relation, formalising which boundary sets cover other sets. This paper works through the possibilities, providing examples to show that all covering relations not immediately ruled out by the definitions are possible.
\end{abstract}

\maketitle
\tableofcontents

%%%%%%%%%%%%%%%%%%%%%%%%%%%%%%%%%%%%%%%%%%%%%%%%%%%%%%%%%%%%%%%%%%%%%%%%%%%%%%%%%%%%%%%%%%%%%%%%%%%%%%%%%%%%%%%%%%%%%%%%%%%%%%%
%%%%%%%%%%%%%%%%%%%%%%%%%%%%%%%%%%%%%%%%%%%%%%%%%%%%%%%%%%%%%%%%%%%%%%%%%%%%%%%%%%%%%%%%%%%%%%%%%%%%%%%%%%%%%%%%%%%%%%%%%%%%%%%

\section{Introduction}

Defining the completion of a manifold with respect to a pseudo-Riemannian metric is much more complicated than the familiar completion of a metric space. The standard mathematical tool for defining boundaries of topological spaces is called a net, and a definition based on nets, taken from \cite{aboundary}, is given in Section \ref{defns}. \\

The singularity theorems of Hawking, Penrose et. all show that in general relativity, one is forced to work with manifolds with nonempty boundary. Understanding the geometry and topology of a space time will therefore reduce to a large extent to the problem of understanding the boundary. There are many different tools for illuminating different aspects of this boundary using the additional structure present on the topological spaces, for example, using causal structures, \cite{causalboundary}, incomplete geodesics, \cite{gboundary}, or Sasaki metrics \cite{bboundary}.\\

Even if true, it is not satisfactory to have to rely on cosmic censorship to be able to determine whether a space-time is singular or not. Anyone who has studied the Schwarzschild solution will be familiar with the concept of a removable singularity.  Informally, this is often described as a singularity that can be removed by a ``change of coordinates''. More precisely, these ``coordinate changes'' are re-embeddings of the pseudo-Riemannian manifold into different Riemannian manifolds with nonequivalent metrics. In order to distinguish between boundary ``points'' that are intrinsically singular, such as curvature or cone singularities, and those that are not, it is necessary to understand the different ways the pseudo-Riemanninan manifold can be enveloped by a larger Riemannian manifold of the same dimension. Re-envelopments of the pseudo-Riemannian manifold give rise to the notion of what boundary sets cover which other boundary sets. This gives an equivalence relationship used to define boundary points. \\

The purpose of this paper is to show by example how the different types of boundary points may cover each other. This is done to a large extent using cut and paste techniques from low dimensional topology.\\

Section \ref{defns} introduces the different types of boundary points, following the classification from \cite{aboundary}. Section \ref{examples} then goes through all the different types of boundary points, and provides examples to show that all covering relations not explicitly ruled out by the definitions are possible.

\subsection*{Acknowledgements} The author would like to thank S. Scott for suggesting the project, and S. Scott and M. Ashley for helpful discussions.

\section{Background and Definitions} 
\label{defns}
The \textit{a}-boundary provides an implicit means for describing and classifying boundary points, and will be outlined briefly here. For more details, see for example \cite{Mikesthesis}, \cite{secondaboundarypaper}, \cite{WAS} \& \cite{aboundary}. 
  
%this paper is about examples.

\begin{defn}[Enveloped manifold]
An \textit{enveloped manifold} is a triple $(\manifold{M}, 
\manifold{\widehat{M}}, \phi)$ where \manifold{M} and 
$\manifold{\widehat{M}}$ are differentiable manifolds of the same 
dimension and $\phi$ is a $C^{\infty}$ embedding $\phi : \manifold{M} 
\rightarrow \manifold{\widehat{M}}$. The enveloped manifold is also 
called an envelopment, where $\manifold{\widehat{M}}$ is the 
enveloping manifold.
\end{defn}

\begin{defn}[Extension]
An \textit{extension} of a pseudo-Riemannian manifold 
(\manifold{M},$g$) is an envelopment of it by a pseudo-Riemannian 
manifold ($\manifold{\widehat{M}}$, $\hat{g}$) such that 
$\hat{g}{\mid}_{\phi(\manifold{M})}=g$.
\end{defn}

Informally speaking, a singularity is an approachable boundary point of \manifold{M}, at which the manifold structure breaks down. In the \textit{a}-boundary \  scheme, approachability is defined with respect to a family of curves satisfying the following property. The set of curves in question is chosen to suit the aim of the investigation.

\begin{defn}[bounded parameter property (b.p.p.)]
A family $\mathcal{C}$ of parametrized curves in \manifold{M} satisfies the 
b.p.p.\ if:
\begin{enumerate}
\item for any point $p$ $\in$ \manifold{M} there is at least one 
curve of the family passing through $p$

\item if $\gamma(t)$ is a curve of the family then so is any 
connected subset of it

\item if $\gamma$ and $\gamma'$ are in $\mathcal{C}$ and $\gamma'$ is obtained from $\gamma$ by a change of parameter then either the parameter is 
bounded or unbounded on both curves.
\end{enumerate}
\end{defn}

Curves satisfying the b.p.p.\ are a generalization of geodesics with 
affine parameter.\\

In order to define the \textit{a}-boundary, it is necessary to make precise 
what is meant by two boundary sets representing the same abstract set in different envelopments. This is done by an equivalence relation.

\begin{defn}[Covering relation]
If $B$ is a boundary set of \phim \ and $B'$ is a boundary set of $\phi^{\prime}(\manifold{M})$ then $B$ \textit{covers} $B'$ (denoted $B \vartriangleright B'$) if for every open neighbourhood \nbhd{U} of $B$ in $\manifold{\widehat{M}}$ there exists an open neighbourhood $\nbhd{U}'$ of $B'$ in $\manifold{\widehat{M}}'$ such that 
\begin{equation}
\phi \circ {\phi'}^{-1}(\nbhd{U}' \cap \phi'(\manifold{M})) 
\subset \nbhd{U}.
\end{equation}
\end{defn}

Two boundary sets $B$ and $B'$ are equivalent if they cover each 
other. This defines an equivalence relation. An equivalence class is 
denoted by a representative element with a square bracket around it, 
for example $[B]$. In the \aboundary \ formulation, a boundary set is an 
equivalence class, and an \aboundary \ point is an equivalence class 
with a point on the boundary of some envelopment as a representative element.

\begin{defn}[Abstract boundary $B(\manifold{M})$]
$B(\manifold{M}):=\{[p]\mid p \in \partial_{\phi}(\manifold{M})$ for 
some envelopment $(\manifold{M}, \manifold{\widehat{M}}, \phi)\}$
\end{defn}

Points of the \aboundary \ are then divided into various categories.

\begin{defn}[Regular point]
A boundary point $p$ of an envelopment (\manifold{M}, $g$, 
$\manifold{\widehat{M}}\text{, }\phi$) is \textit{regular} if there exists a 
manifold ($\manifold{\overline{M}}, \overline{g}$) such that 
$\phi(\manifold{M}) \cup \{p\} \subseteq \manifold{\overline{M}} 
\subseteq \manifold{\widehat{M}}$ and $(\manifold{M}, g, 
\manifold{\overline{M}}, \overline{g}, \phi)$ is an extension of 
(\manifold{M}, $g$).
\end{defn}

Regularity does not ``pass to the \aboundary'' (i.e.\ it is not 
invariant under the equivalence relation used to define the \aboundary). A regular \aboundary \ point is defined as follows:

\begin{defn}[Regular \aboundary \ point]
A regular \aboundary \ point is an equivalence class with a regular 
point as a representative element.
\end{defn}

A maximally extended space-time is essentially one whose \aboundary\  
does not contain any regular \aboundary\  points. 

\begin{defn}[Maximally extended]
A $C^k$ pseudo-Riemannian manifold (\manifold{M}, $g$) is termed 
$C^l$ maximally extended $(1 \leqslant l \leqslant k)$ if there does 
not exist a $C^l$ extension (\manifold{M}, $g, 
\widehat{\manifold{M}}, \widehat{g}, \phi$) of (\manifold{M}, $g$) 
such that $\phi(\manifold{M})$ is a proper open submanifold of 
$\widehat{\manifold{M}}$
\end{defn}

The Misner example \cite{HawkingEllis}, is a simplification of an example given in \cite{Taubnut}. One description of it is the cylinder $\mathbb{R}\times S^{1}$, with the metric
\begin{equation*}
ds^{2}=-2dtd\theta+td\theta^{2}
\end{equation*}
The Misner example has a closed null curve at $t=0$, and this closed curve is approached by null geodesics that are incomplete with respect to their affine parameters. However, the submanifold $t=0$ would appear to be perfectly regular, interior points of the manifold. In the \aboundary, irregularity replaces the concept of curve 
incompleteness often used in the literature as an indicator of the existence of singularities. This has the advantage of being able to distinguish between examples such as the Misner example, in which the existence of incomplete geodesics are not really indicative of singular behaviour.
%in the other boundary constructions. The two concepts 
%are very closely related, however. Scott and Ashley \cite{Mike's 
%thesis} have proven a result linking curve incompleteness with the 
%existence of \aboundary \  essential singularities, under various 
%conditions. Results like these demonstrate that the rule of 
%thumb that ``an incomplete curve is an indication of singular 
%behaviour'', which occurs throughout the literature, gives a result 
%in agreement with the \aboundary \ under very general conditions. 

\begin{defn}[Limit point of a curve]
We say that $p \in \phim\cup\bphim$ is a \textit{limit point} of a curve $\gamma :[a,b)\rightarrow \phim$ if there exists an increasing infinite sequence of real numbers $t_i \rightarrow b$ such that $\gamma(t_i)\rightarrow p$.
\end{defn}

\begin{defn}[Endpoint of a curve]
We say that $p$ is an \textit{endpoint} of the curve $\gamma$ if $\gamma(t)\rightarrow p$ as $t\rightarrow b$.
\end{defn}

\begin{defn}[Approachable boundary point]
A parametrised curve $\gamma :I \rightarrow \manifold{M}$ \textit{approaches} the boundary set $B$ if the curve $\phi \circ \gamma$ has a limit point lying in $B$. A point $p\in \bphim$ is \textit{approachable} if it is approached by a curve from the family $\mathcal{C}$. 
\end{defn}

Irregular boundary points consist of singularities, points at infinity and irregular unapproachable boundary points.

\begin{defn}[Point at infinity]
A boundary point $p$ of the envelopment (\manifold{M}, $g, 
\widehat{\manifold{M}}, \mathcal{C}, \phi$) is a \textit{point at infinity} if
\begin{enumerate}
\item $p$ is not a regular boundary point
\item $p$ is approachable by an element of $\mathcal{C}$, and
\item no curve of $\mathcal{C}$ approaches $p$ with bounded parameter.
\end{enumerate}
\end{defn}

\begin{defn}[Removable point at infinity]
A boundary point $p$ at infinity is termed a \textit{removable point at infinity} if there is a boundary set, $B\subset\bphim$ composed purely of regular boundary points such that $B\triangleright p$
\end{defn}

\begin{defn}[Essential point at infinity]
A point $p$ at infinity is an \textit{essential point at infinity} if it is not removable.
\end{defn}

\begin{defn}[Mixed point at infinity]
An essential point $p$ at infinity is a \textit{mixed point at infinity} if it covers a regular boundary point.
\end{defn}

\begin{defn}[Pure point at infinity]
An essential point at infinity is a \textit{pure point at infinity} if it does not cover any regular boundary points.
\end{defn}

\begin{defn}[Singular boundary points]
A boundary point $p$ of an envelopment (\manifold{M}, $g, 
\widehat{\manifold{M}}, \mathcal{C}, \phi$) is called \textit{singular} or a 
\textit{singularity} if 
\begin{enumerate}
\item $p$ is not a regular boundary point,
\item $p$ is approachable by a curve $\gamma$, where $\gamma$ is 
an element of $\mathcal{C}$ and has finite parameter.
\end{enumerate}
\end{defn} 

\begin{defn}[Removable singularity]
A singular boundary point $p$ will be called \textit{removable} if it can be covered by a non-singular boundary set $B$ of another embedding. 
\end{defn}

\begin{defn}[Essential singularity]
A singular boundary point $p$ is called \textit{essential} if it is 
not removable.
\end{defn}

\begin{defn}[Directional and pure singularities]
An essential singularity $p$ is called a \textit{directional 
singularity} if it covers a boundary point of another embedding which is either regular or 
a point at infinity. Otherwise $p$ is called a \textit{pure 
singularity}.
\end{defn}

The definitions of directional and pure singularities pass to the 
\aboundary, as shown in \cite{aboundary}.\\

\begin{figure}
\centering
\includegraphics{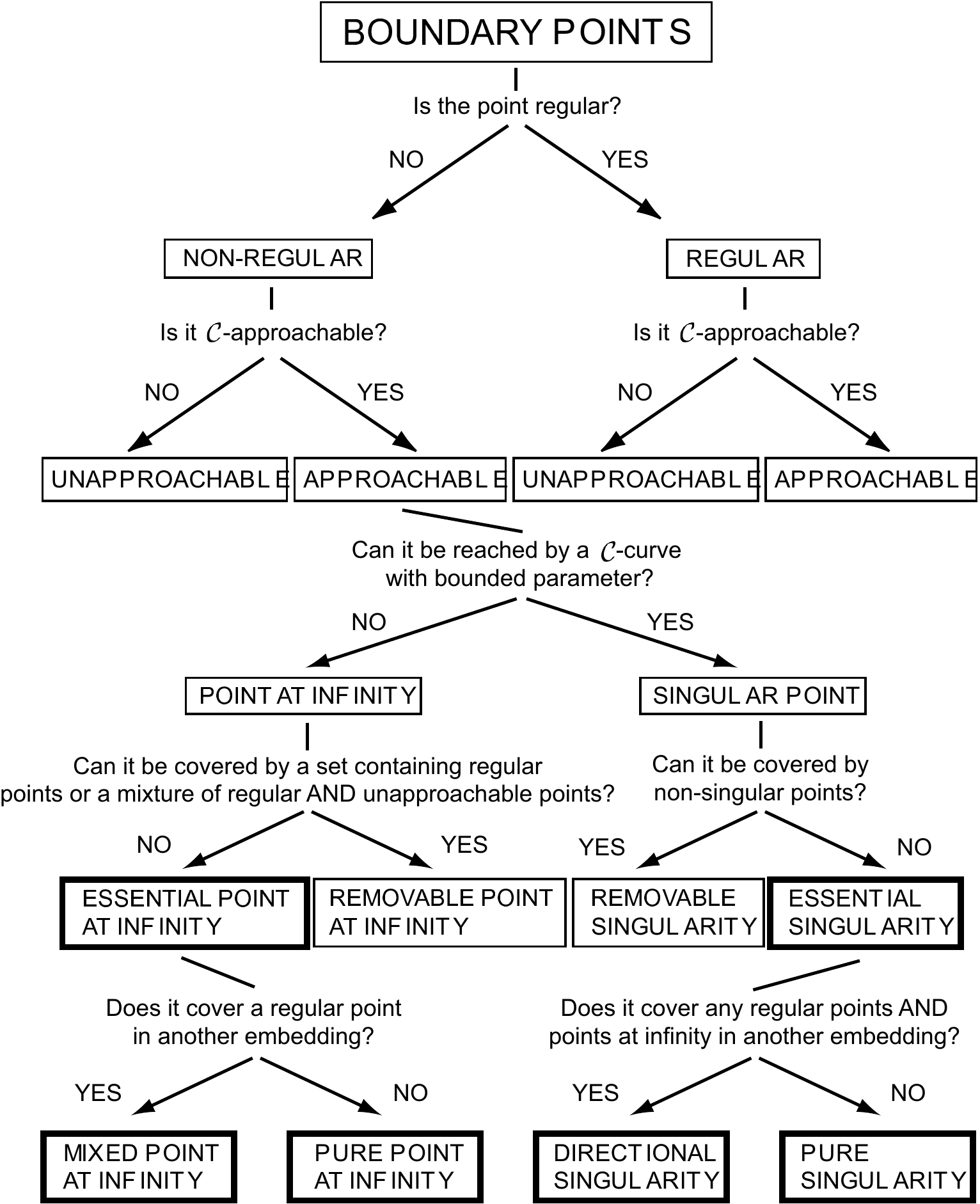}
\caption{A figure taken from \cite{aboundary} summarising the \aboundary\ point classification scheme. Categories that pass to the \aboundary \ are in bold.}
\label{a-boundary flowchart}
\end{figure}

The \aboundary classification scheme is summarised in Figure \ref{a-boundary flowchart}.

A concept that has turned out to be useful is that of a connected neighbourhood region, \cite{Mikesthesis}.

\begin{defn}[Connected Neighbourhood Region (CNR)]
Suppose $p \in \partial_{\phi}\manifold{M}$ and \nbhd{N} is a 
neighbourhood of $p$ in \mhat. Then a connected component of 
$\nbhd{N} \cap \phim$ is called a \textit{connected neighbourhood 
region of} $p$. 

\end{defn}
\begin{figure}
\centering
\includegraphics[height=0.15\textheight]{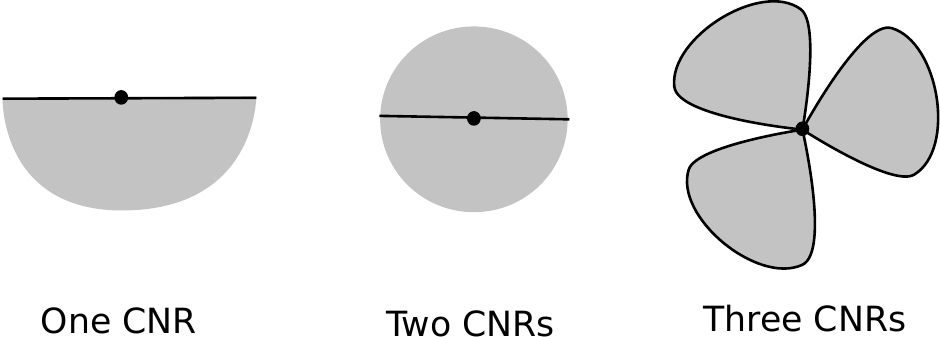}
\caption{In this diagram, \phim \ is represented by the shaded 
region. This figure gives examples of different numbers of connected neighbourhood regions of the boundary point indicated by the black dot.}
\end{figure}

\begin{defn}[The finite connected neighbourhood region 
property (FCNR property)]
We say that $p$ has $n$ \textit{connected neighbourhood regions} if 
for any open neighbourhood $\nbhd{N}(p)$ there exists a 
sub-neighbourhood $\nbhd{U}(p) \subset \nbhd{N}(p)$ for which 
$\nbhd{U}(p)\cap\phim$ is composed of exactly $n$ connected 
components, and $n$ is the smallest natural number for which this is 
true. The boundary point $p$ satisfies the \textit{finite connected 
neighbourhood region property} if it has only finitely many 
connected neighbourhood regions.
\end{defn}

It is a consequence of Theorem 4.3 of \cite{secondaboundarypaper} 
that the FCNR property passes to the \aboundary .

\section{Examples}
\label{examples}
\begin{figure}[h!]
\centering
\includegraphics[height=0.30\textheight]{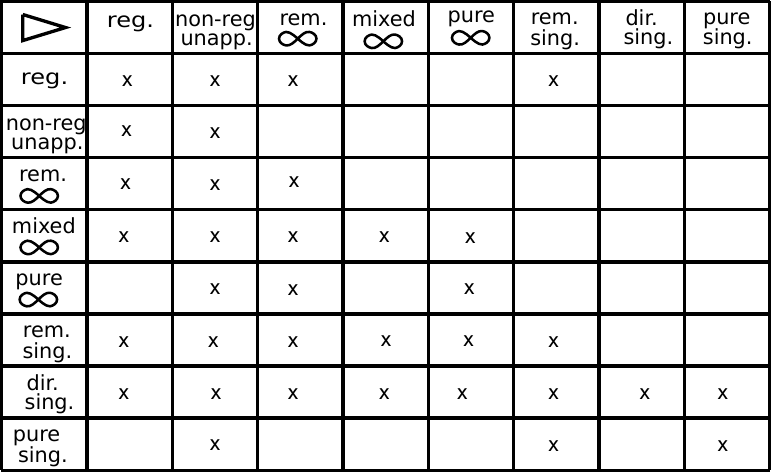}
\caption{A table summarising the types of boundary points that may and may not cover each other. An x implies the entry in the column can cover the corresponding entry in the row.}
\label{coveringtable}
\end{figure}

Figure \ref{coveringtable} is a table displaying which types of  \aboundary \ points can cover other types. When it is not possible for a particular type of boundary point to cover another, this is an immediate consequence of the definitions. In all other cases, the entries in the table will now be confirmed by providing examples. \\

%\begin{figure}[h!]
%\centering
%\includegraphics[height=0.30\textheight]{figures/coveringtable.pdf}
%\caption{A table summarising the types of boundary points that may and may not cover each other. This Figure is taken from \cite{aboundary}.}
%\label{coveringtable}
%\end{figure}

In this section, it will be assumed that $\mathcal{C}$ is the set of geodesics with affine parameter. \\

\textbf{Separating out a point.} When a boundary point $p$ has a finite number $k$ of connected 
neighbourhood regions, many of the examples presented in this section will be constructed by splitting boundary points up into points with fewer connected neighbourhood regions, and recombining these in different ways. The procedure outlined below for splitting up a boundary point with $k$ connected neighbourhood regions into $k$ boundary points, each with one connected neighbourhood region, will be referred to as ``separating out a point''.\\

Suppose \phim \ is an $n$ dimensional manifold and $\nbhd{N}_1$ is a connected neighbourhood region of the point $p\in \bphim$ with $k>1$ connected neighbourhood regions, $\nbhd{N}_{1}, \ldots, \nbhd{N}_k$. Let $S \subseteq \mhat \backslash \phim$ be a closed codimension one surface with $p$ in its interior. The surface $S$ will also be referred to as a slit, and may have nonempty boundary in \mhat. Suppose also that there is some neighbourhood $\mathcal{N}$ of $p$ in which $S$ is separating, and in this neighbourhood $\nbhd{N}_1$ is on one side of $S$ and the other connected neighbourhood regions are on the other. More precisely, $S$ is chosen so that for every open neighbourhood \nbhd{U} of $p$ in \mhat, there exists an open neighbourhood $\nbhd{V} \subseteq \nbhd{U}$ of $p$ in \mhat, such that $\nbhd{V} \setminus S$ has two connected components in \mhat, $\nbhd{V}_1$ and $\nbhd{V}_2$, where $\nbhd{N}_1$ has nonempty intersection with precisely one of the components, say $\nbhd{V}_1$, and $\{\nbhd{N}_{2}, \ldots,\nbhd{N}_k\}$ have non-empty intersection only with $\nbhd{V}_2$. Since $p$ is a boundary point with more that one connected neighbourhood region and \mhat \  is a manifold, a set $S$ with these properties can always be found. \\

\begin{figure}
\centering
\includegraphics[height=0.10\textheight]{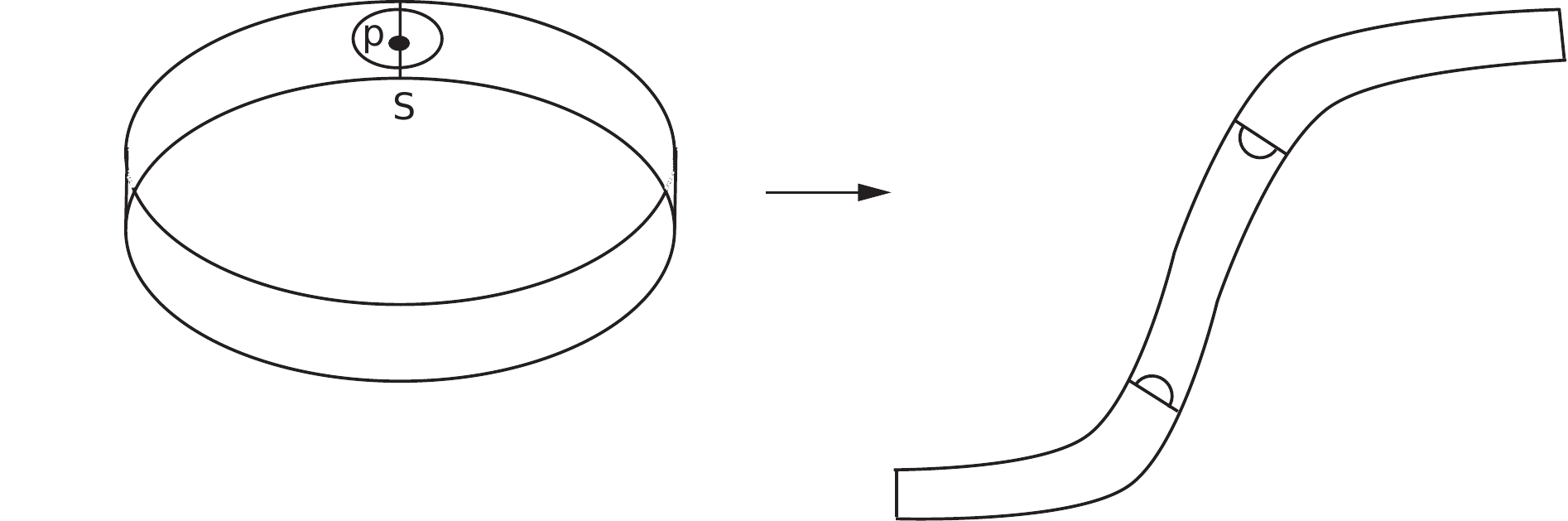}
\caption{The two neighbourhood regions at $p$ are separated out.}
\label{separateout}
\end{figure}

Remove $S$ from \mhat \ and  identify the lower edge of the slit with the upper edge of the slit in a second copy of $\mhat 
\backslash S$, as shown in Figure \ref{separateout}. Identify the upper edge of the slit with the lower edge of the slit in a third copy of $\mhat \backslash S$. In this envelopment, $p$ is equivalent to two boundary points, one with only one connected neighbourhood region, and the other with $k-1$ connected neighbourhood regions. If $k>2$ this process is repeated on the boundary point with more than one connected neighbourhood region, until a single enveloping manifold is obtained, in which $p$ is equivalent to $k$ boundary points each with only one connected neighbourhood region.\\

If $S$ can be chosen in such a way that it does not have any boundary points in common with \phim, as will always be the case in this paper, then separating out a point does not destroy the regularity properties of any \aboundary \ points, including $p$. Separating out a point also does not change the original embedding $\phi$, it merely enlarges the enveloping manifold, by taking the union of $\mhat \setminus S$ with two copies $\mhat_1$ and $\mhat_2$ of $\mhat \setminus S$. The union is joined along the copies of $S$ as explained above.

\begin{ex}[A non-regular unapproachable boundary point that covers a regular 
point]
\label{example1}
Put coordinates (r,$\theta$) on ${\thereals}^2$ and let
\manifold{M} be the manifold satisfying $r>1$, 
$r\cos\theta<1\text{, }0<\theta<2\pi$ with metric given by
\begin{equation*}
ds^2=\frac{-1}{\theta}dr^2 + r^2d{\theta}^2.
\end{equation*}

\begin{figure}
\centering
\includegraphics[height=0.15\textheight]{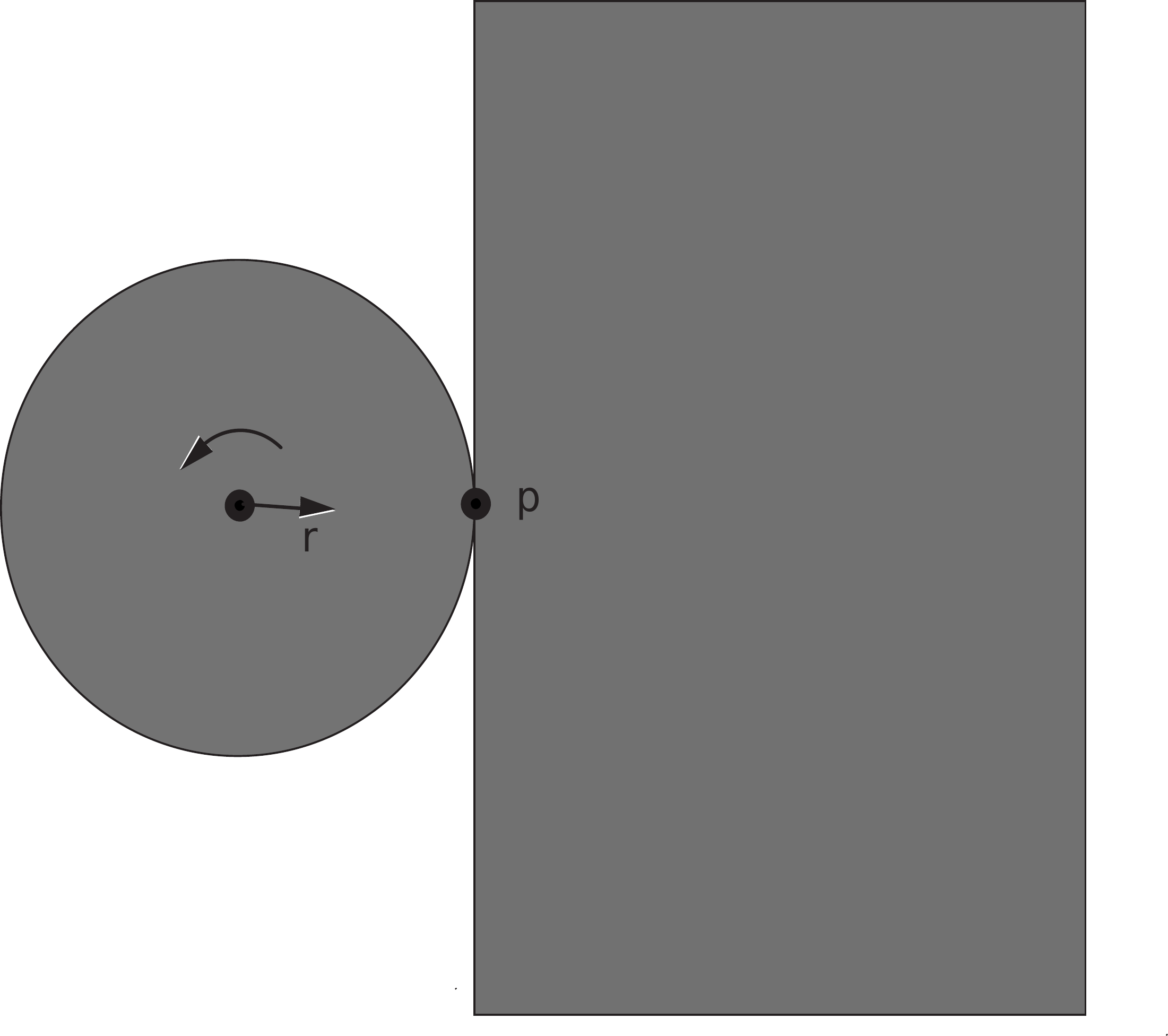}
\caption{In this figure, the manifold is $\phi(\mathcal{M})$ is the unshaded region. The radial coordinates have their origin at the dot in the center of the circle.}
\label{figure1a}
\end{figure}

Let $p$ be the boundary point with coordinates $r=1$ and $\theta =0$, as shown in Figure \ref{figure1a}. 
Then $p$ is non-regular. The curvature scalar is given by
\begin{equation*}
\frac{1}{r}-\frac{3}{2r^{2}\theta^{2}}+\frac{2\theta}{r^{2}}
\end{equation*}
Curves approaching $p=(1, 2\pi)$ with increasing $\theta$ have a finite limit of the curvature scalar, while curves
approaching $p=(1, 0)$ with decreasing $\theta$ have unbounded curvature
scalar. The removal of the circle from the space-time makes $p$
unapproachable. Separating out the two connected neighbourhood regions reveals a regular point $(1,2\pi)$ and a non-regular $(1,0)$ point covered by $p$.
\end{ex}

\begin{ex}[A removable point at infinity that covers a regular point]
\label{spike}
Put coordinates $(x,y)$ on ${\thereals}^2$ and let \manifold{M} be the manifold satisfying $y<-1$, 
 $\mid x \mid < e^y$, with metric
\begin{equation*}
ds^2=-dy^2 + dx^2
\end{equation*}
The re-envelopment $\phi^{\prime}$ of \manifold{M} in ${\thereals}^2$ given by
\begin{equation*}
x \rightarrow x'=x\text{, } y \rightarrow y'=\arctan(y).
\end{equation*}
has precisely one boundary point which is a point at infinity, namely the point $p$, with coordinates 
 $(x',y')=(0, -\pi/2)$. 
To see that $p$ is a removable point at infinity, consider a second re-envelopment $\phi^{\prime \prime}$ of \manifold{M} in ${\thereals}^2$ given by
\begin{equation*}
x\rightarrow x''=x+4e^{-1}-4e^y\text{, }y\rightarrow y''=y\text{\  mod 1}.
\end{equation*}
The action of this re-envelopment is depicted in Figure \ref{windup}. The image $\phi''(\manifold{M})$ is contained in the compact subset 
\begin{equation*}
\{(x'',y'')|-e^{-1}\leqslant x''\leqslant 4e^{-1}, 0\leqslant y'' \leqslant 1\} 
\end{equation*}
of $\mhat''$. The point $p$ is equivalent to the set of regular points
\begin{equation*}
 \{(x'',y'')|x''= 4e^{-1}, 0\leqslant y'' \leqslant 1\}.
\end{equation*}
Therefore $p$ is removable and covers a regular boundary point.
\end{ex}

\begin{figure}
\centering
\includegraphics[height=0.30\textheight]{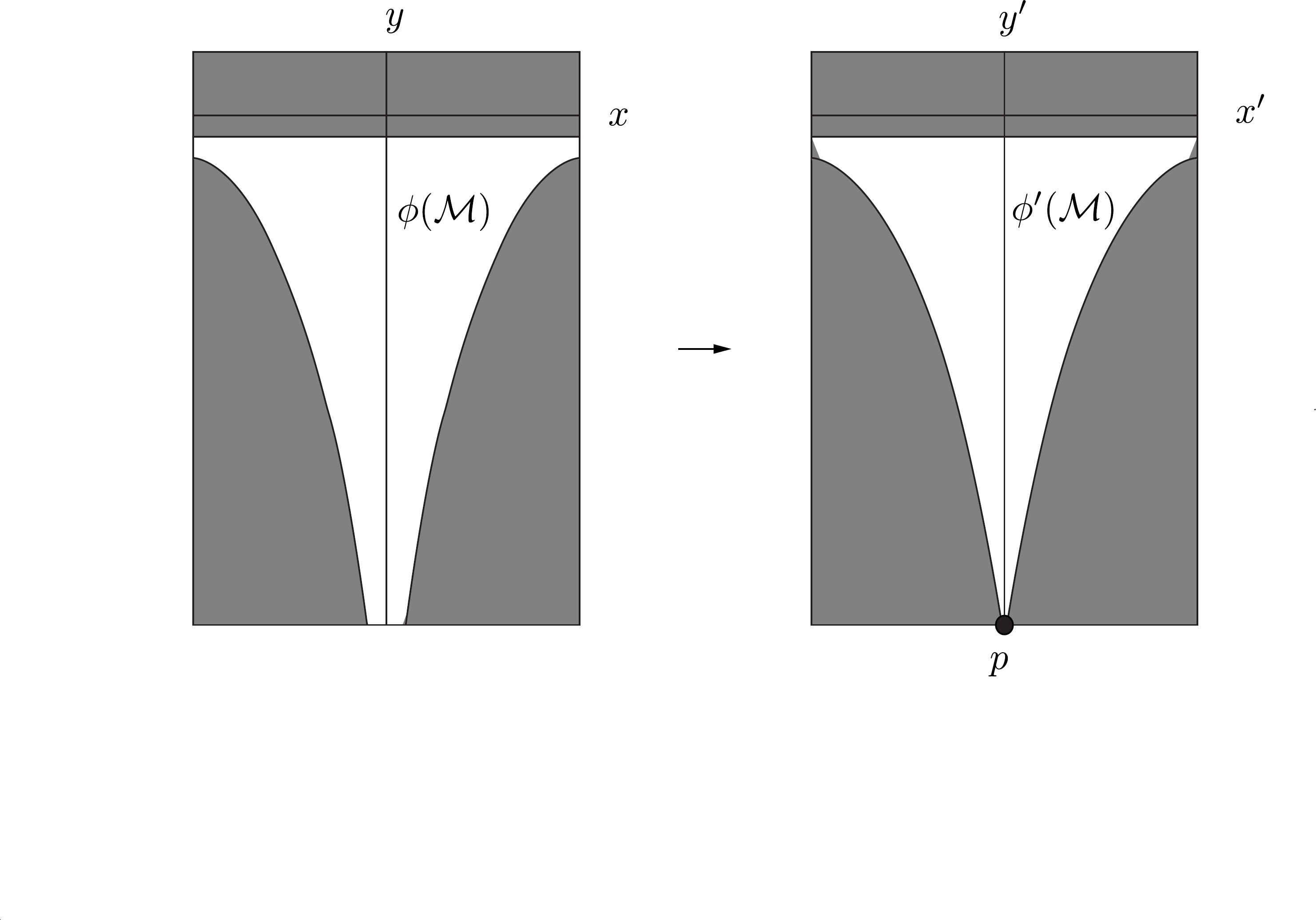}
\caption{The point $p$ is a removable point at infinity.}
\label{figure2}
\end{figure}

\begin{figure}
\centering
\includegraphics[height=0.30\textheight]{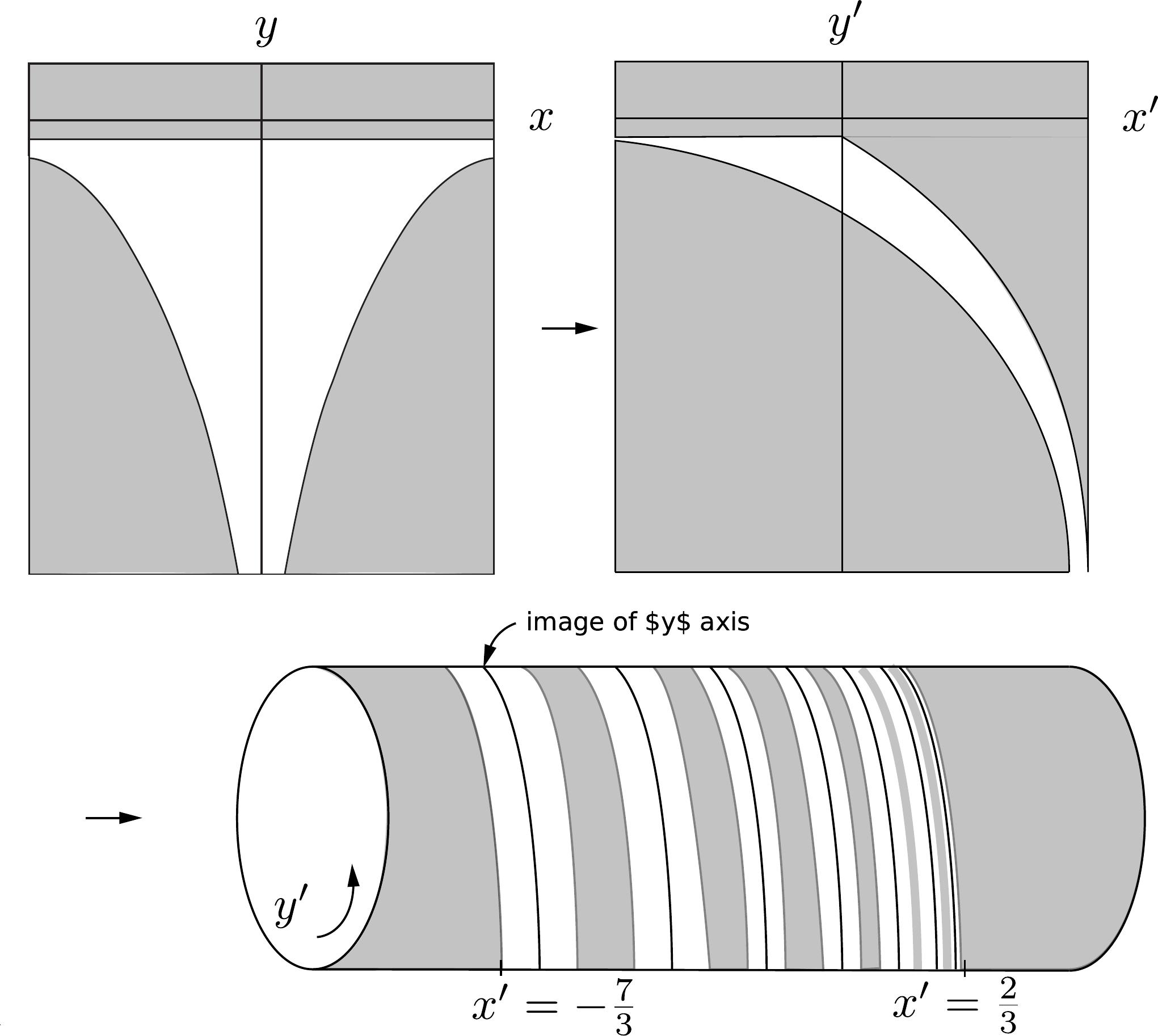}
\caption{The construction from Example \ref{spike}.}
\label{windup}
\end{figure}

\begin{figure}
\centering
\includegraphics[height=0.15\textheight]{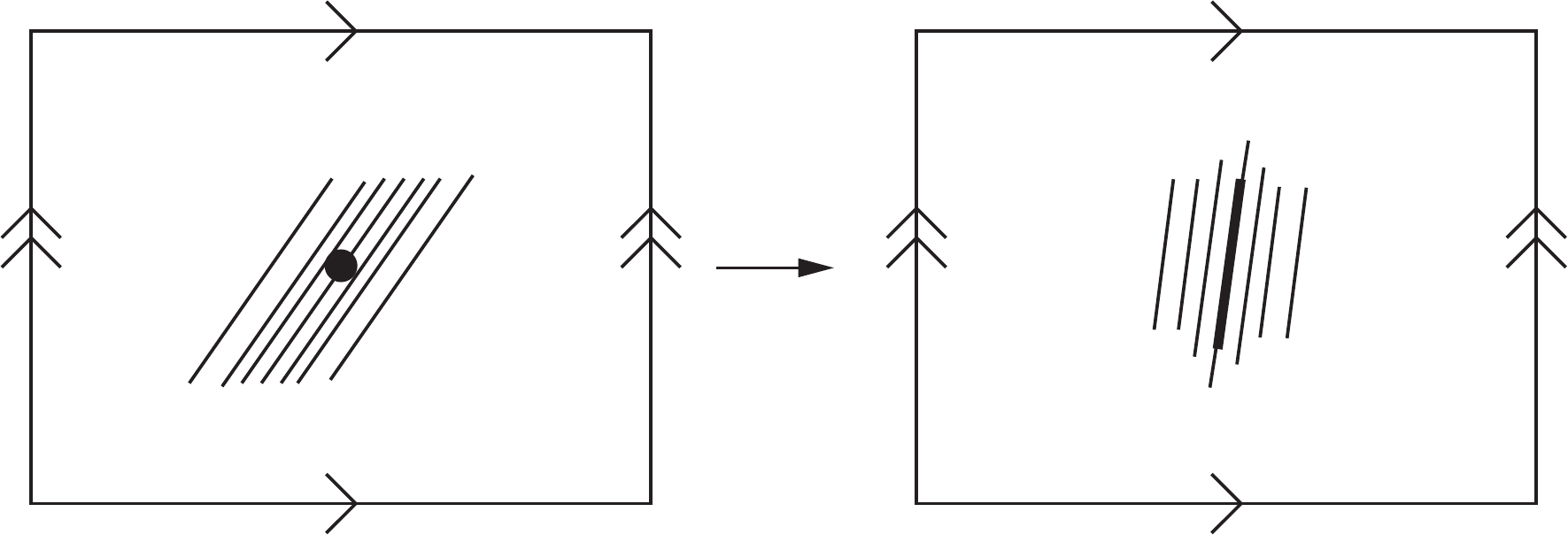}
\caption{The Figure from Example \ref{donutexample}.}
\label{donut}
\end{figure}

\begin{ex}[A regular point that covers a removable point at infinity]
\label{donutexample}
Example 33 of \cite{aboundary} contains a regular boundary point that is only approachable by geodesics with infinite affine parameter. In this example, \mhat \ is the unit torus with metric $ds^2 = dx^2+dy^2$. On the central line $L=\{(x,1/2)|0\leqslant x<1\}$ choose the points 
\begin{equation*}
p_{\pm i}=\left(\frac{1}{2}\left(1\pm\frac{1}{2^i}\right),\frac{1}{2}\right),
\hspace{1.5cm}i=1,2,3,\ldots
\end{equation*}
For each $i=\pm1,\pm2,\ldots$ let $L_i$ be the closed line segment of length $1/2$ and slope $\sqrt{2}$ centered on the point $p_i$ and let $L_0$ be a similar line segment with center $p$. Here \phim \ is the open submanifold of \mhat\ consisting of the complement in \mhat \ of this infinite set of closed line segments, as shown in Figure \ref{donut}. The point $p$ is clearly a regular boundary point and is approached only by infinite geodesics with $\frac{dy}{dx}=\sqrt{2}$. The re-envelopment
\begin{equation*}
x\rightarrow x'=x\text{, }y\rightarrow y'=\arctan \left(\frac{y}{x^2} \right)
\end{equation*} 
blows up the point $p$ into a compact interval $I$ of irregular boundary points. By theorem 19 of \cite{aboundary}, $[p]=[I]$, therefore there is an irregular point covered by $p$ that is only approachable by infinite geodesics, i.e. $p$ covers a point at infinity.
\end{ex}

%Example 39 of \cite{aboundary} shows a removable singularity
%covering a regular point, Example 45 shows a directional
%singularity covering a regular point, and Example 25 shows a
%regular boundary point covering an irregular unapproachable point.

\begin{ex}[A removable singularity covering a regular point, (Example 39 of \cite{aboundary})]
Consider the manifold $\mathbb{R}\setminus (0,0)$ with the metric 
\begin{equation*}
ds^{2}=dr^{2}+(r+1)^{2}d\theta^{2}
\end{equation*}
The boundary point $p=(0,0)$ is approachable but not regular. One way of seeing that it is not regular is to observe that for circles centered on $p$, the ratio of circumference to radius does not approach $2\pi$ but approaches infinity as the radius approaches zero. The coordinates $(r, \theta)$  therefore do not provide a coordinate patch that can be extended to $(0,0)$, hence $p$ is not regular.\\

Take the re-envelopment determined by
\begin{equation*}
\phi'(x,y)=\frac{r+1}{r}(r\cos\theta, r\sin\theta)
\end{equation*}
Now $\phi'(\manifold{M})$ is the region $r>1$ of $\mathbb{R}^{2}$, with the flat metric 
\begin{equation*}
ds^{2}=dr^{2}+r^{2}d\theta^{2}
\end{equation*}
The boundary can now be seen to consist of a set of regular boundary points, equivalent to $p$. The point $p$ is therefore a removable singularity that covers regular points.
\end{ex}

\begin{ex}[A directional singularity covering a regular point]
Example 45 of \cite{aboundary} studies such an example in detail, however the metric is slightly complicated. An easier way to construct an example of a directional singularity covering a regular point is to take the previous example, and use a partition of unity to multiply the flat metric on $\phi'(\manifold{M})$ by a conformal factor. This is done in such a way that one and only one of the boundary points of $\phi'(\manifold{M})$ becomes a curvature singularity, while the others remain regular. Then take the re-embedding $\phi\circ \phi^{'-1}$ to obtain a directional singularity covering a regular point.
\end{ex}

\begin{ex}[A regular boundary point covering an irregular unapproachable point]
Let $\manifold{M}$ be the manifold $\mathbb{R}^{2}\setminus (0,0)$. Blowing up the point $p=(0,0)$, i.e. taking the re-envelopment

\begin{gather*}
(x', y')=
\begin{cases}
(\arctan\left( \frac{x}{y}\right), \sqrt{x^{2}+y^{2}}) \text{ for } 0\leq x,\text{ and }0\leq y\\
(\arctan\left( \frac{x}{y}\right)+\pi, \sqrt{x^{2}+y^{2}})\text{ for } x<0,\\
(\arctan\left( \frac{x}{y}\right)+\pi, \sqrt{x^{2}+y^{2}})\text{ otherwise}
\end{cases}
\end{gather*}
gives a re-embedding $\phi'(\manifold{M})\simeq S^{1}\times \mathbb{R}^{+}$. In this re-envelopment, $[p]$ has representative the ring $y'=0$. Each of the boundary points of $\phi'(\manifold{M})$ is approached by just one geodesic contained in $\phi'(\manifold{M})$. The boundary points of $\phi'(\manifold{M})$ can therefore not be regular. This is because for regular points, the inverse function theorem implies the exponential map is a local diffeomorphism. In the same way, blowing up any of the boundary points of $\phi'(\manifold{M})$ give irregular unapproachable points. These points are covered by the regular point $p$.
\end{ex}

\begin{figure}
\centering
\includegraphics[height=0.25\textheight]{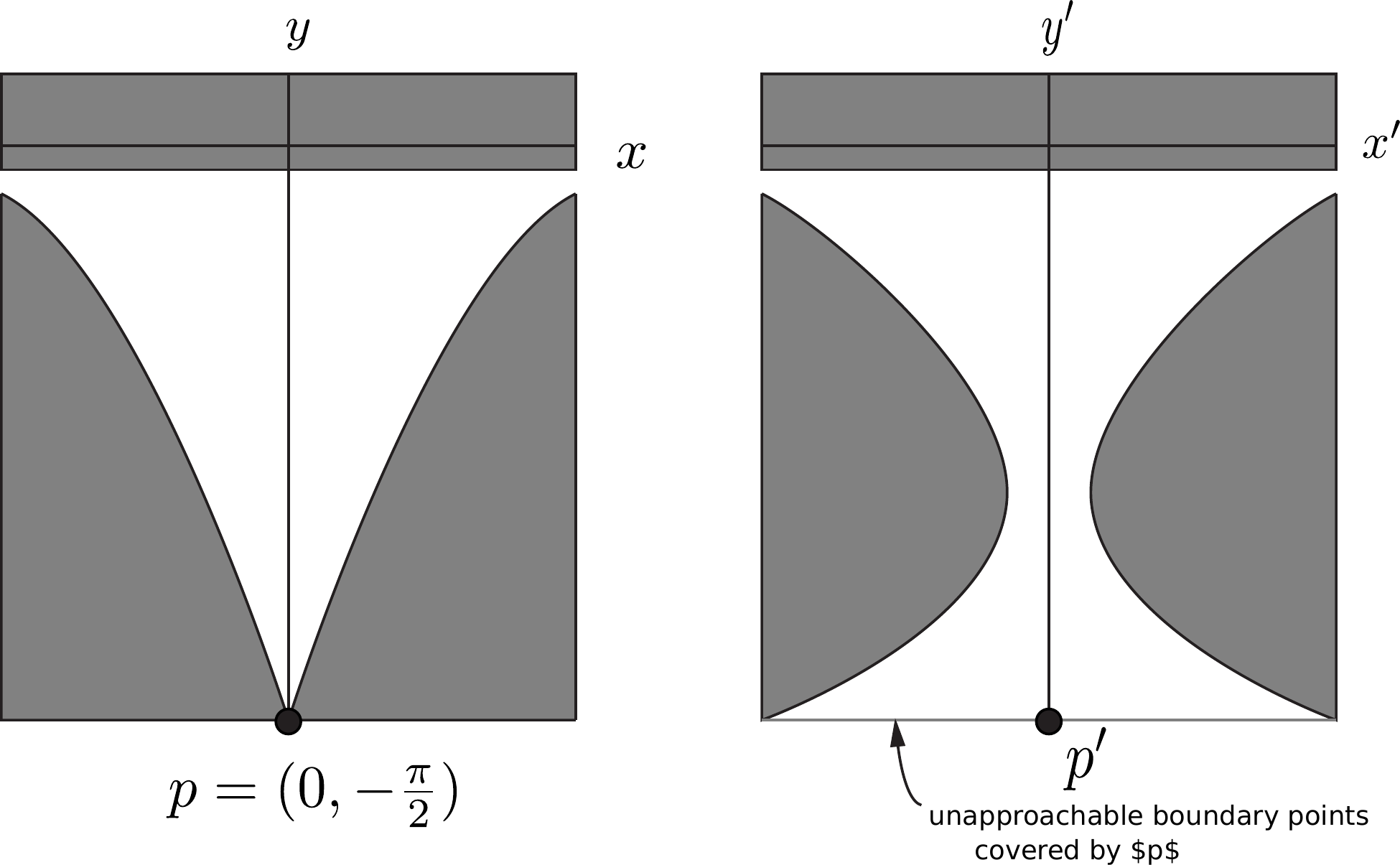}
\caption{The Figure from example \ref{bluntspike}.}
\label{figure3}
\end{figure}

\begin{ex}[A removable point at infinity covering an irregular 
unapproachable point]
\label{bluntspike}
Consider the removable point at infinity from Example \ref{spike}. The width
$w(y')$ of the manifold around the $y'$ axis is given by
\begin{equation*}
w(y')=\frac{2}{y^{\prime 4}}.
\end{equation*}
The envelopment $x'\rightarrow x''= x'e^{\frac{1}{w(y')}}\text{, }y'\rightarrow y''=y'$ reveals unapproachable irregular boundary
points covered by $p$, see Figure \ref{figure3}.
\end{ex}

%4

\begin{figure}
\centering
\includegraphics[height=0.25\textheight]{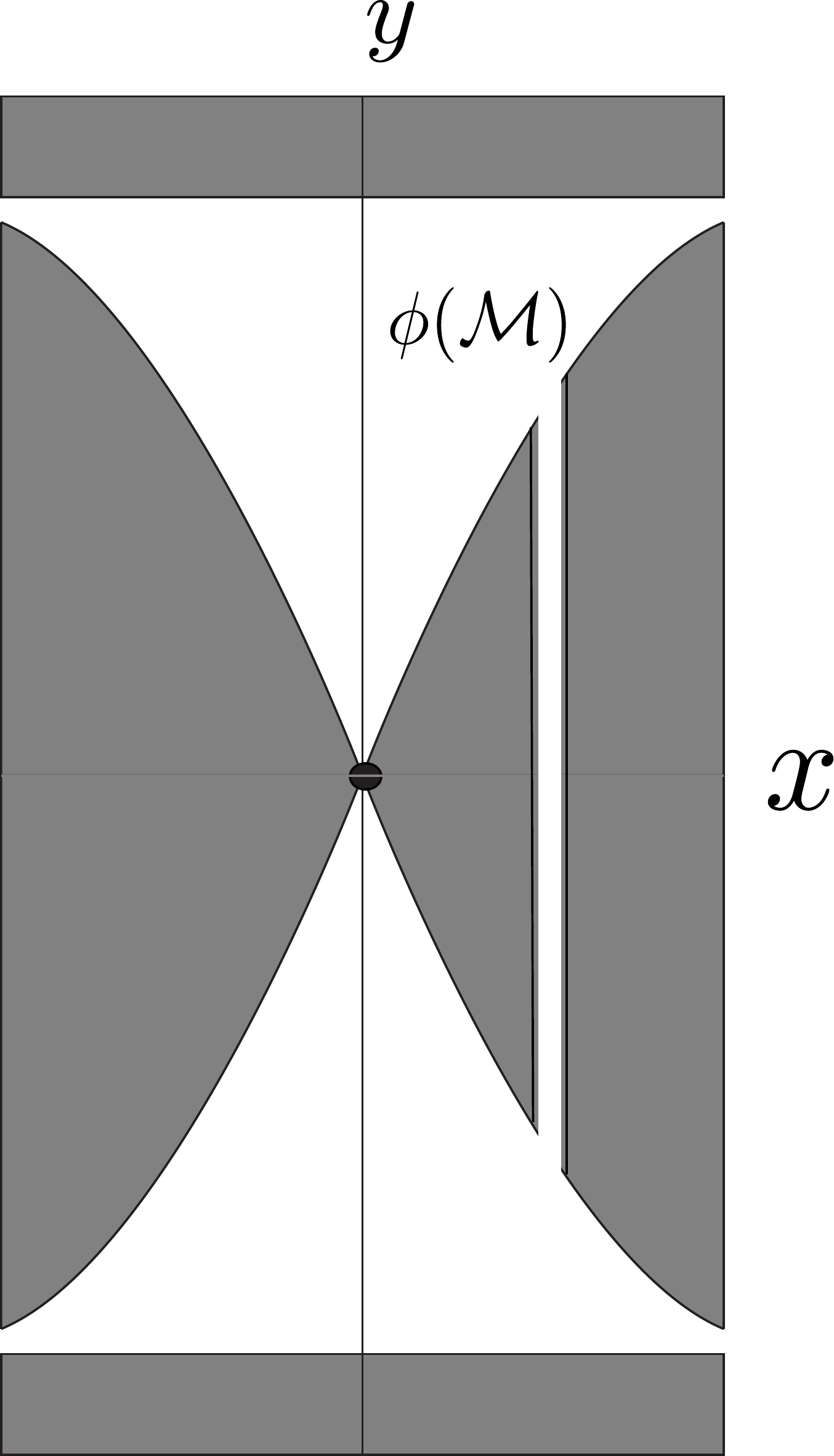}
\caption{The Figure from example \ref{doublespike}.}
\label{figure7}
\end{figure}

\begin{ex}[A mixed point at infinity covering a removable point at infinity and an irregular unapproachable point]
\label{doublespike}
Start with the manifold containing a point at infinity given in
example \ref{spike}. Multiplying the metric by $\frac{1}{(x^{\prime 2}+(y+\pi/2)^{\prime 2})^2}$ makes the point $(0,-\pi/2)$ a pure point at infinity. Attach the manifold from example \ref{spike}, as shown in Figure \ref{figure7}. \\

Let $g_{1}$ be the metric on the upper half of the diagram, and let $g_{2}$ be the metric on the lower part of the diagram. In order to make this manifold connected, attach a ``bridge'' from the top part to the lower part, as shown in the diagram.
Let $l$ be a
parameter along the connecting piece, such that $l=0$ on the boundary of the upper part of the example, and increases smoothly to $l=1$ on the boundary of the other part. Then \manifold{M} is the
manifold consisting of the two components + ``bridge'', where the
metric $g$ is given by
\begin{equation*}
g|_{\text{region1}}=g_1\text{, }g|_{\text{region2}}=g_2 \text{ and }  g|_{\text{bridge}}=(1-l)g'_1 + lg'_2
\end{equation*}
The two metrics are simultaneously diagonalizable, so it is easy to verify that the metric obtained in this way is nonsingular everywhere on \manifold{M}.\\

The origin is a mixed point at infinity, since it is not removable, but covers the removable point at infinity from example \ref{spike} that covers regular points. As shown in the previous example, this removable point at infinity covers irregular unapproachable points. 
\end{ex}

\begin{ex}[A pure point at infinity covering an irregular 
unapproachable point]
This example is the same as example \ref{bluntspike} only the removable point at infinity is made into a pure point at infinity by multiplying the metric by the conformal factor $\frac{1}{(x^{\prime 2}+(y+\pi/2)^{\prime 2})^2}$.
\end{ex}

\begin{ex}[A removable singularity covering an irregular 
unapproachable point]

Start with the region $y>0$ of two dimensional space with
metric $ds^2=-dy^2+dx^2$. The re-envelopment
\begin{equation*}
x \rightarrow x'=\frac{x}{y^2}\text{, }y \rightarrow y'=y
\end{equation*}
makes the origin, $p$, in these new coordinates a removable
singularity, approached only by the geodesic $x'=0$.
Repeating this process (only with $x'$ instead of $x$ and
$y'$ instead of $y$) reveals irregular unapproachable points
covered by $p$.
\end{ex}
%7
\begin{ex}[A directional singularity covering an irregular 
unapproachable point and a removable singularity]

Example 45 of \cite{aboundary} is a directional singularity. The details of this example are not required here.
To show that it covers an irregular unapproachable point, choose a
regular point $p$ covered by the singularity. 
Put normal coordinates $(x,y)$ around $p$. The re-envelopment
\begin{equation*}
x \rightarrow x'=\frac{x}{y^2}\text{, }y \rightarrow y'=y
\end{equation*}
fixes the geodesic locally given by x=0, and sends all the
other geodesics approaching $p$ off to infinity. Any point other
than the origin on the $x'$ axis is an irregular unapproachable
point covered by the directional singularity. The origin of the
$x'$ axis is a removable singularity covered by a directional
singularity.
\end{ex}

Example 25 of \cite{aboundary} is a pure singularity (To be
more specific it is a cone singularity, as defined in \cite{EllisandSchmidt}) which covers irregular unapproachable points.

\begin{ex}[A pure singularity covering a removable singularity, and a pure point at infinity covering a removable point at infinity]

Start with the two dimensional manifold with metric
\begin{equation*}
ds^2=\frac{1}{\sqrt{x^2+y^2}}(-dy^2+dx^2), \hspace{1.5cm} \frac{\pi}{4} <
\theta < \frac{3\pi}{4}.
\end{equation*}
Then take two dimensional manifold
\begin{equation*}
\{(x,y)|  y>0\}\text{, with metric } ds^2=-dy^2+dx^2
\end{equation*}
and re-envelop it as follows:
\begin{equation*}
x \rightarrow x'=\arctan \left(\frac{x}{y}\right)\text{, } y \rightarrow y'=y
\end{equation*}
The point $(x', y')=(0,0)$ is a removable singularity. The manifold
\manifold{M} formed by identifying the origins of these two
manifolds and inserting a connecting piece has a pure singularity
at the origin which covers a removable singularity.\\

\begin{figure}
\centering
\includegraphics[height=0.20\textheight]{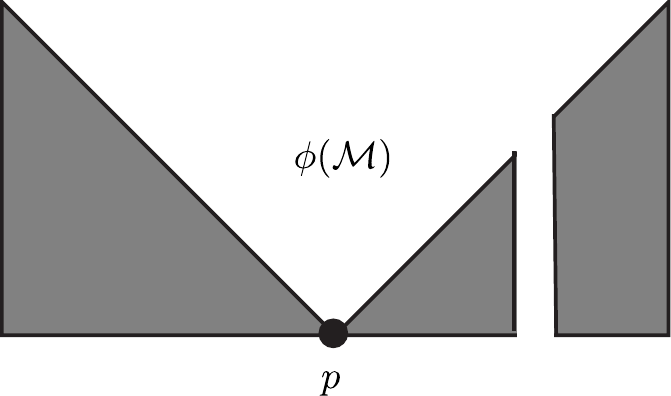}
\caption{A pure singularity covering a removable singularity.}
\label{figure8}
\end{figure}

An example of a pure point at infinity covering a removable point at 
infinity is formed from the previous example by sending the singularity off to infinity.
\end{ex}

\begin{figure}
\centering
\includegraphics[height=0.20\textheight]{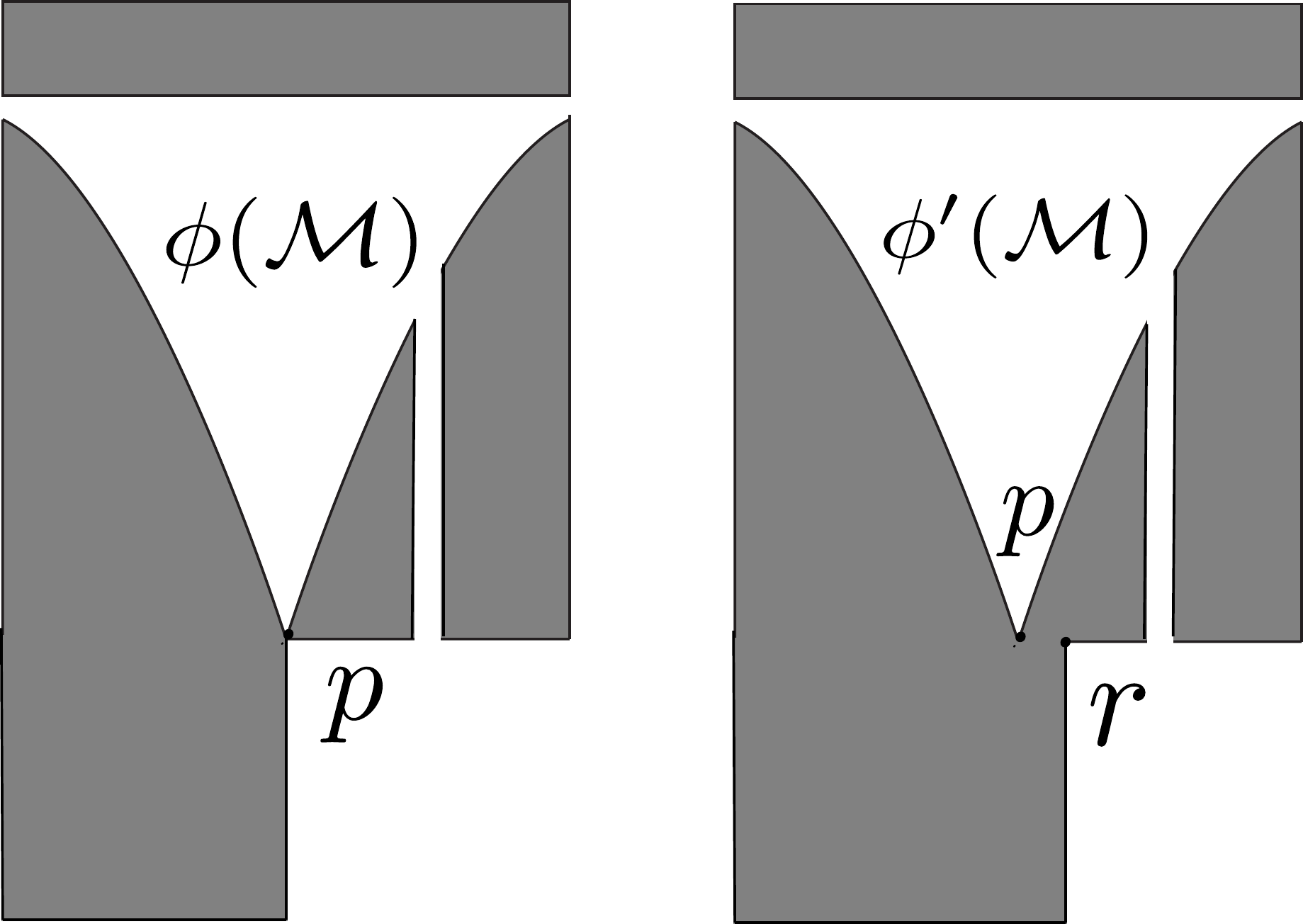}
\caption{A removable singularity covering a removable point at 
infinity.}
\label{figure9}
\end{figure}

\begin{ex}[A removable singularity covering a removable point at
infinity]

\manifold{M} consists of the manifold in example \ref{spike} containing a removable point at infinity, connected to a quadrant of flat space, with a connecting piece, as shown in Figure \ref{figure9}. The point $p$ is singular because it is irregular and approachable by finite geodesics. By theorem 19 of \cite{aboundary} it is equivalent to the \aboundary \ set consisting of $[q]$ - a removable point at infinity, and $[r]$ - a regular point. Therefore $p$ is a removable singularity which covers a removable point at infinity.
\end{ex}

\begin{figure}
\centering
\includegraphics[height=0.25\textheight]{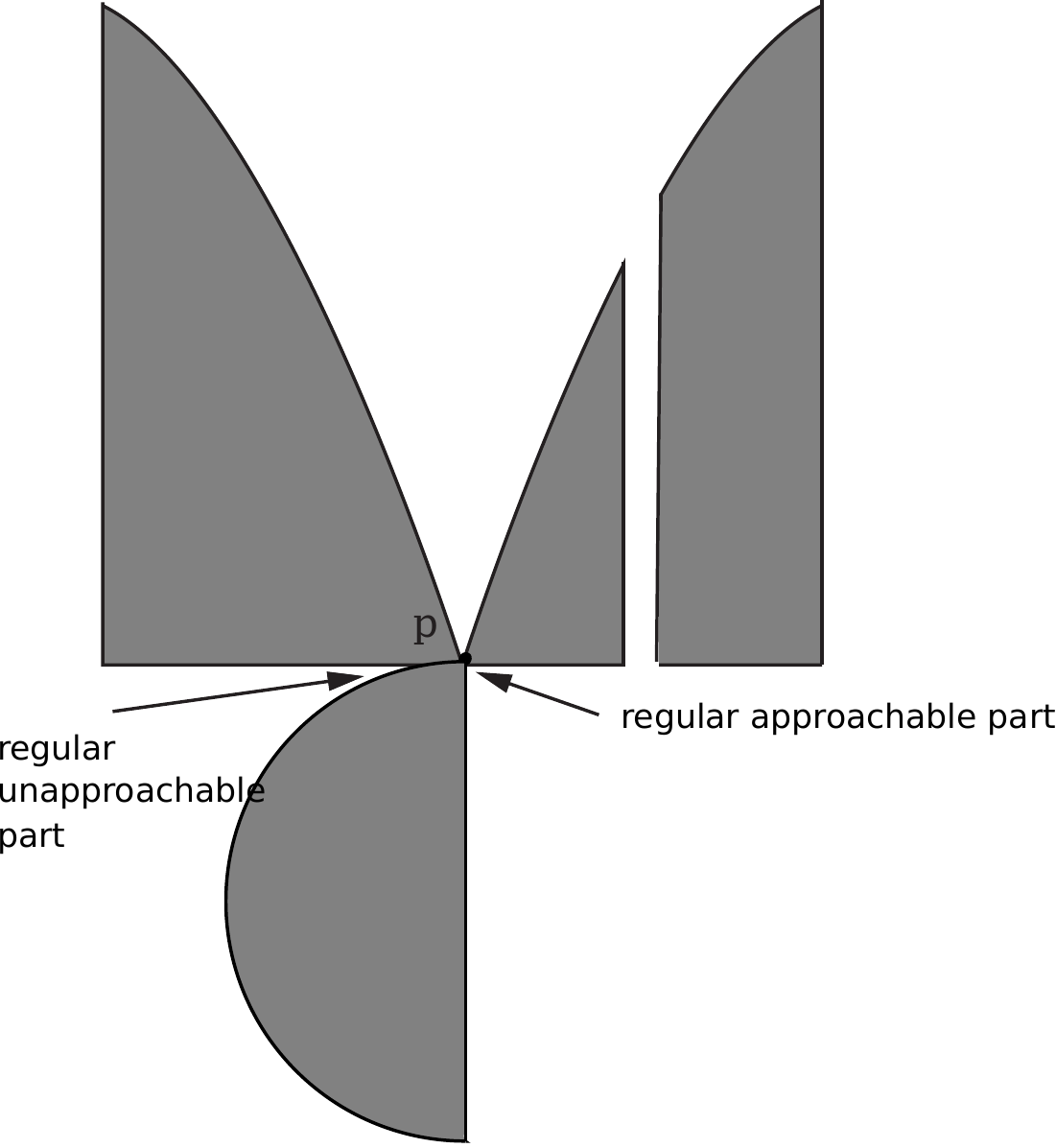}
\caption{A removable singularity covering a mixed point at infinity 
and a pure point at infinity.}
\label{figure10}
\end{figure}

\begin{ex}[A directional singularity covering a removable point at
infinity]

 This example is the same as the previous example,
except that the metric on the quadrant is replaced by
\begin{equation*}
ds^2=\frac{1}{\sqrt{x^2+y^2}}(-dy^2+dx^2).
\end{equation*}
This turns $p$ into a directional singularity but does not
influence the classification of the point $q$ in the
re-envelopment.
\end{ex}

\begin{ex}[A removable singularity covering a mixed point at
infinity and a pure point at infinity]

Consider the manifold
\begin{equation*}
\{(x,y)|y<0\}\text{ with metric }ds^2=dx^2+dy^2.
\end{equation*}
Remove a semicircle from this space as shown in Figure \ref{figure10}. This is done in such a way that the boundary
point $(0,0)$ has two connected neighbourhood regions, one
associated with a regular unapproachable point and the other
associated with a regular point approached by finite geodesics.
Take the top half of example \ref{doublespike} and identify the pure point at infinity with the point $(0,0)$ of the manifold just constructed, and call this boundary point $p$. Make this manifold connected as in the earlier examples. The point $p$ is irregular, approachable by geodesics with finite affine parameter, and is equivalent to a set consisting of regular points and points at infinity. Therefore $p$ is a removable singularity. Separating off the regular approachable part of $p$ reveals a mixed point at infinity covered by $p$. This mixed point at infinity covers the pure point at infinity from Example \ref{doublespike}.
\end{ex}

%13
\begin{ex}[A directional singularity covering a mixed point at
infinity and a pure point at infinity]
This is the same as the previous example except with the following alteration. Break the point $p$ up into its connected neighbourhood regions. Let $q$ be the regular approachable point covered by $p$. Using a partition of unity, alter the metric around the point $q$ by a conformal factor so that $q$ becomes a curvature singularity. Putting the three connected neighbourhood regions back together again results in a directional singularity that covers a mixed point at infinity (i.e. the boundary point with the neighbourhood region around which the metric has been altered removed) and a pure point at infinity.
\end{ex}

The Curzon 
solution \cite{Curzon1} \& \cite{Curzon2} contains an example of a directional singularity that covers a pure singularity.

\small
\bibliographystyle{alpha}
%\nocite{*}
\bibliography{gravitybib}

\begin{thebibliography}{{Mis}67}

\bibitem[Ash02]{Mikesthesis}
M.~Ashley.
\newblock {\em Singularity theorems and the abstract boundary construction}.
\newblock PhD thesis, Australian National University, 2002.

\bibitem[ES79]{EllisandSchmidt}
G.~Ellis and B.~Schmidt.
\newblock Classification of singular space-times.
\newblock {\em General Relativity and Gravitation}, 10(12):989--997, Aug 1979.

\bibitem[FHS11]{causalboundary}
J.~L. Flores, J.~Herrera, and M.~Sanchez.
\newblock {On the final definition of the causal boundary and its relation with
  the conformal boundary}.
\newblock {\em Adv. Theor. Math. Phys.}, 15(4):991--1057, 2011.

\bibitem[FS94]{secondaboundarypaper}
C.~Fama and S.~Scott.
\newblock {Invariance properties of boundary sets of open embeddings of
  manifolds and their application to the abstract boundary}.
\newblock {\em Submitted to: Contemp. Math.}, 1994.

\bibitem[Ger68]{gboundary}
R.~Geroch.
\newblock Local characterization of singularities in general relativity.
\newblock {\em Journal of Mathematical Physics}, 9(3):450--465, 1968.

\bibitem[HE73]{HawkingEllis}
S.~W. Hawking and G.~F.~R. Ellis.
\newblock {\em The Large Scale Structure of Space-Time}.
\newblock Cambridge Monographs on Mathematical Physics. Cambridge University
  Press, 1973.

\bibitem[{Mis}67]{Taubnut}
C.~{Misner}.
\newblock {\em {Taub-Nut Space as a Counterexample to almost anything}}.
\newblock 1967.

\bibitem[Sch71]{bboundary}
B.~G. Schmidt.
\newblock A new definition of singular points in general relativity.
\newblock {\em General Relativity and Gravitation}, 1(3):269--280, Sep 1971.

\bibitem[SS86a]{Curzon1}
S.~Scott and P.~Szekeres.
\newblock The curzon singularity. {I}: Spatial sections.
\newblock {\em General Relativity and Gravitation}, 18(6):557--570, Jun 1986.

\bibitem[SS86b]{Curzon2}
S.~Scott and P.~Szekeres.
\newblock The curzon singularity. {II}: Global picture.
\newblock {\em General Relativity and Gravitation}, 18(6):571--583, Jun 1986.

\bibitem[SS94]{aboundary}
S.~Scott and P.~Szekeres.
\newblock {The Abstract boundary: A new approach to singularities of
  manifolds}.
\newblock {\em J. Geom. Phys.}, 13:223--253, 1994.

\bibitem[WAS15]{WAS}
B.~Whale, M.~Ashley, and S.~Scott.
\newblock Generalizations of the abstract boundary singularity theorem.
\newblock 32, 07 2015.

\end{thebibliography}

\end{document}